\documentclass[%
 reprint,
 amsmath,amssymb,
pra,
]{revtex4-1}
\usepackage{graphicx}
\usepackage{dcolumn}
\usepackage{bm}
\newcommand{\be}{\begin{equation}}
\newcommand{\ee}{\end{equation}}
\newcommand{\bea}{\begin{eqnarray}}
\newcommand{\eea}{\end{eqnarray}}
\newcommand{\se}{Schr\"odinger equation }

\begin{document}
\title{On the evolution of quantum non-equilibrium in expanding systems}
\author{Samuel Colin}
\email{samuel.colin@u-cergy.fr}
\affiliation{%
Laboratoire de Physique Th\'eorique et Mod\'elisation, CNRS Unit\'e 8089,  
	CY Cergy Paris Universit\'e, 95302 Cergy-Pontoise cedex, France}
\date{\today}
\begin{abstract}
We consider a particle confined in a uniformly expanding two-dimensional square box from the point of the view of the de Broglie-Bohm pilot-wave theory.
In particular we study quantum ensembles in which the Born Law is initially violated (quantum non-equilibrium). 
We show examples of such ensembles that start close to quantum equilibrium, as measured by the standard coarse-grained H-function, but diverge from it  with time.
We give an explanation of this result and discuss the possibilities that it opens.
\end{abstract}
\maketitle
\section{Introduction}
According to standard quantum theory, there is no reality in the quantum world, outside of measurements, and quantum particles are meant to be entirely described by wave-functions. 
In the alternative theory called the de Broglie-Bohm pilot-wave theory \cite{debroglie,bohm1,bohm2}, the wavefunction is not the ultimate description of a particle, or of our universe. 
It is merely a non-local agent orchestrating the motion of supplementary elements of reality, elements whose existence is independent from measurements, thereby avoiding many paradoxes plaguing standard
quantum theory. 
In the case of a single non-relativistic particle, this element of reality (or configuration) is simply the position of the
particle and it evolves in a deterministic way under the guidance of the wave-function. 
The existence of these preferred configurations has two deep implications which we outline in the next two paragraphs.

In the context of quantum cosmology, the so-called quantum minisuperspace models, which are simplified models of the universe characterized by a few (say N) degrees of
freedom, can be unambiguously studied within the pilot-wave approach \cite{npnjfa}. Actually their analysis is similar to that of a non-relativistic particle in
a N-dimensional configuration space, the main difference being that the wave-function is now a solution of the Wheeler-DeWitt 
equation, instead of being a solution of the Schroedinger equation. Given the deterministic character of the pilot-wave theory, it is sufficient
to specify the initial condition of the preferred configuration and a solution of the Wheeler-DeWitt equation in order to obtain the entire trajectory for
the model universe, and to assert, for example, whether the universe can undergo a bounce \cite{mnsepb} for certain sets of models. 

The second implication of the existence of preferred configurations is linked to the fact that they do not need to be distributed according to
the Born law in an ensemble -- by ensemble we have in mind a set of copies of the same quantum system (at different spacetime events) described by the same wave-function.
If they are distributed in that way, the predictions of standard quantum theory are reproduced, and the ensemble
is said to be in quantum equilibrium but if they are not, the ensemble is in quantum non-equilibrium, and standard quantum theory is
violated \cite{ava1,ava2,avaphd}. 
Thanks to numerical simulations, ensembles in quantum non-equilibrium have been shown to relax quickly to quantum equilibrium, for wavefunctions obtained by superposing many energy modes, yielding sufficiently complex dynamics \cite{vawe,toruva}.
For wavefunctions containing low numbers of modes, relaxation would generally not be complete and the ensemble would have a residual quantum non-equilibrium \cite{abcova}. 
The more energy eigenmodes the wave-function contains, the less probable it is to find some residual quantum non-equilibrium in the ensemble.
Correspondingly the amount of residual quantum non-equilibrium, if there is some, is expected to decrease as a function of the number of modes contained in the wave-function \cite{abcova}.
Finally quantum equilibrium is a terminal state: if an ensemble relaxes to quantum equilibrium, it will stay there --
thus it is unlike a medium in classical equilibrium which could perturbed and set in non-equilibrium.
All these facts are consistent with a cosmological scenario in which the universe started in a state of primordial
quantum non-equilibrium, but quickly relaxed to quantum-equilibrium thanks to its violent history. If this is indeed the case, quantum nonequilibrium
could still manifest itself today, either if primordial quantum non-equilibrium left some imprints in the cosmic microwave
background \cite{ava3,cova1,cova2,vipeva}, or through the existence of relic quantum non-equilibrium particules \cite{avajpa,unva1,unva2} -- which are particles that would be decoupled at very early times, 
while still in quantum non-equilibrium and whose wave-function is simple enough to prevent total relaxation to quantum equilibrium.

The closeness of a non-equilibrium distribution with respect to quantum equilibrium is measured by the so-called coarse-grained H-function \cite{ava1}.
All previous studies showed that if an ensemble is close to quantum equilibrium, it won't move further away from quantum equilibrium.
However, all these studies, apart from the case of a scalar field on a de Sitter space \cite{cova1}, dealt with non-expanding sytems. 
Here we show that it is generally not true for a system in expansion, consisting of a particle trapped in a two-dimensional square box whose walls are set in uniform motion.

The article is organized as follows. In section \ref{sec:sqm} we generalize the solutions for a particle in a uniformly expanding one-dimensional box  to the two-dimensional case. 
Then we introduce two definitions (modes and number of modes) relevant for the class of wave-functions that we consider.
In section \ref{sec:pwt} we apply the pilot-wave aproach to those wave-functions; we give the explicit form of the guidance equations and we illustrate the kinds of dynamics that it predicts. 
We recall the notion of quantum non-equilibrium and how to measure the difference with respect to quantum equilibrium thanks to the coarse-grained H-function.
Then we show three examples leading to an increase in the coarse-grained H-function, two of them starting relatively close to quantum equilibrium.
Finally we explain why this increase is occuring.
The article ends with section \ref{sec:dis} in which we discuss the possibilities opened by this line of research.
\section{Standard Quantum Mechanics}\label{sec:sqm}
\subsection{Solutions}
We consider a particle of mass $m$ confined in an expanding two-dimensional square box whose side length at time $t$ is given by $L(t)=L_0+v_e t$. 
The two spatial coordinates are denoted by $x_1$ and $x_2$. 
The potential is zero for $(x_1,x_2)\in[0,L(t)]\times[0,L(t)]$ and infinite elsewhere.

The interior wave-function $\psi(t,x_1,x_2)$ must satisfy the \se 
\be
i\hbar\partial_t\psi(t,x_1,x_2)=-\frac{\hbar^2}{2m}(\partial^2_{x_1}+\partial^2_{x_2})\psi(t,x_1,x_2)\label{se}
\ee
together with the time-dependent boundary conditions
\bea
&\psi(t,0,x_2)=\psi(t,L(t),x_2)=0\nonumber\,,&\\
&\psi(t,x_1,0)=\psi(t,x_1,L(t))=0\,.&
\eea

The solutions to the one-dimensional version of this problem were first obtained in \cite{dori}.
They were rederived as part of a larger program in \cite{made}; there the authors used a change of variables which led to an equivalent system with time-independent boundary conditions.

It is straightforward to generalize the last method to two dimensions. We introduce the coordinates 
$y_{k}=x_k/s(t)$, with $s(t)=L(t)/L_0$ and $k\in\{1,2\}$, and the function $\tilde{\psi}(t,y_1,y_2)=\psi(t,x_1,x_2)=\tilde\psi(t,\frac{x_1}{s(t)},\frac{x_2}{s(t)})$. 
Then Eq. (\ref{se}) becomes
\bea
i\hbar\partial_t\tilde\psi(t,y_1,y_2)=-\frac{\hbar^2}{2m s^2(t)}(\partial^2_{y_1}+\partial^2_{y_2})\tilde\psi(t,y_1,y_2)\nonumber\\
+i\hbar\frac{\dot{s}(t)}{s(t)}(y_1\partial_{y_1}\tilde\psi(t,y_1,y_2)+y_2\partial_{y_2}\tilde\psi(t,y_1,y_2))\,.\label{se2}
\eea
If we define 
\be
\tilde{\psi}(t,y_1,y_2)=\frac{1}{s(t)}e^{i\alpha(t)\frac{{y_1}^2+{y_2}^2}{2}}\tilde\phi(t,y_1,y_2)\,,
\ee
with $\alpha=\frac{m\dot{s}s}{\hbar}$, then Eq. (\ref{se2}) becomes
\be
i\hbar\partial_t\tilde\phi(t,y_1,y_2)=-\frac{\hbar^2}{2 m s^2}(\partial^2_{y_1}+\partial^2_{y_2})\tilde\phi(t,y_1,y_2)\,.
\ee
A final change change of variables $\tau=t/s(t)$ and the introduction of $\phi(\tau,y_1,y_2)=\tilde\phi(t,y_1,y_2)$ lead to
\be
i\hbar\frac{\partial\phi}{\partial\tau}=-\frac{\hbar^2}{2 m}(\partial^2_{y_1}+\partial^2_{y_2})\phi(\tau,y_1,y_2)\,.
\ee
The problem therefore amounts to that of a particle inside an infinite 2D static square well of side length $L_0$, whose solutions are given 
by
\be
\phi_{n_1 n_2}(\tau,y_1,y_2)=\frac{2}{L_0} e^{-i\frac{E_{n_1 n_2}\tau}{\hbar}}\sin{\frac{n_1\pi y_1}{L_0}}\sin{\frac{n_2\pi y_2}{L_0}}\,,
\ee
with $E_{n_1 n_2}=\frac{\hbar^2\pi^2(n^2_1+n^2_2)}{2mL^2_0}$.
The solutions in the first coordinate system read
\bea
\psi_{n_1,n_2}=\frac{1}{2L}e^{-i\frac{E_{n_1 n_2}\tau(t)}{\hbar}}e^{i\frac{m v}{2\hbar L}(x_1^2+x_2^2)}\times\nonumber\\
\sin{\frac{n_1\pi x_1}{L}}\sin{\frac{n_2\pi x_2}{L}}\label{mode}
\eea
They are orthonormal and we will call them ``modes''.
\subsection{Definition of the number of modes}
We define an order among modes in the following way: the mode $k_1 k_2$ precedes the mode $l_1 l_2$ if $E_{k_1 k_2}<E_{l_1 l_2}$.
We will consider solutions of the \se obtained by superposing the first $N$ modes.
The modes will be equally weighted in the superposition but will be multiplied by random phases. 
The generic form of the wave-functions under consideration will therefore be
\be
\psi(t,x_1,x_2)=\sum_{n=1}^{n=N}\frac{e^{i\phi(n)}}{\sqrt{N}}\psi_{n_1(n),n_2(n)}(t,x_1,x_2)\,.\label{wf}
\ee
In case of ambiguity, the highest modes will be randomly chosen. 

Here is an example for $N=9$. The ordered modes are
\begin{align}
11\nonumber\\
12,21\nonumber\\
13,22,31\nonumber\\
14,23,32,41\nonumber\\
\ldots
\end{align}
To build a superposition with $N=9$, we would have to pick (randomly) three combinations among $\{14,23,32,41\}$ and superpose them to the first six modes.
The wave-function that we will consider is of the form $N=10$ (so no ambiguity) and its coefficients can be found in the appendix.

\section{Pilot-Wave Theory}\label{sec:pwt}
\subsection{Equations of motion}
In the pilot-wave theory, an element of a quantum ensemble is not only described by its wave-function $\psi(t,{\vec{x}})$ but also by an actual position $\vec{x}(t)=(x_1(t),x_2(t))$. The wave-function always evolves according to the \se 
whereas the position evolves according to the guidance equation
\be
{\vec{v}}(t)=\frac{\hbar}{m}\mathfrak{Im}\left(\frac{{\vec\nabla}\psi}{\psi}\right)\bigg|_{\vec{x}=\vec{x}(t)}\,.
\ee
The distribution of particle positions over an ensemble is denoted by $\rho(t,\vec{x})$. 
It can be equal to $|\psi(t,{\vec{x}})|^2$, in which case the predictions of standard quantum mechanics are reproduced, but that is not mandatory.
The equations of motion ensure  that $\rho(t,{\vec{x}})=|\psi(t,{\vec{x}})|^2$, provided $\rho(t_i,{\vec{x}})=|\psi(t_i,{\vec{x}})|^2$ for some earlier time $t_i$.

\begin{figure}
\centering
\includegraphics[width=0.45\textwidth]{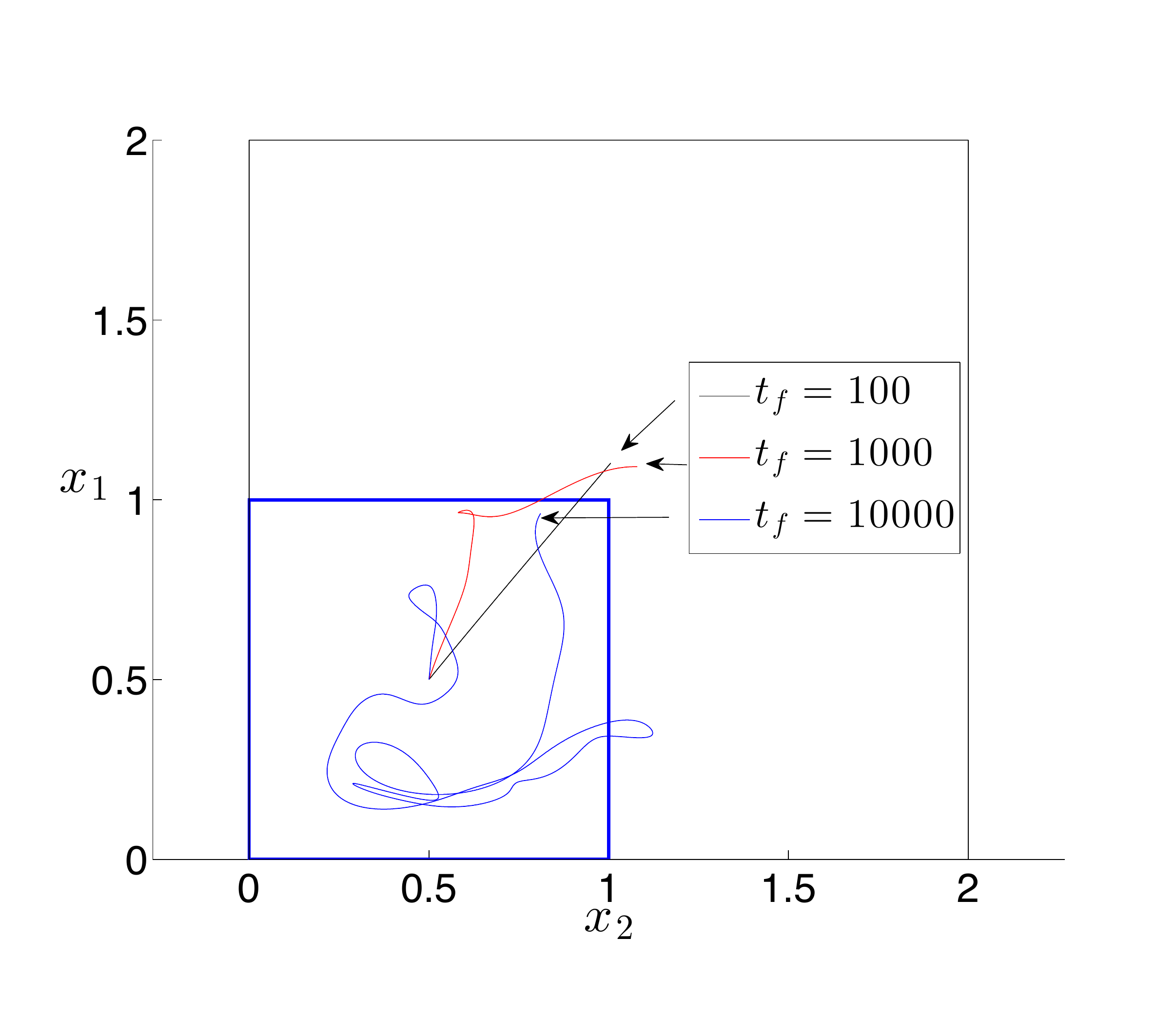}
\caption{\label{fig:fig1}Three trajectories corresponding to three different experiments. 
The initial condition is the same for all three cases but the velocity of expansion and the final times are different.
However the quantity $v_{e}t_f$ is the same for the three cases and is equal to $1$. 
At $t=t_f$ the surface of the box has quadrupled.}
\end{figure}

The uniformly expanding one-dimensional cavity has already been considered from the point of view of the pilot-wave theory (see \cite{mousavi} and \cite{mamowa} for example).
In the  two-dimensional case, the explicit expression for the velocity field is
\begin{widetext} 
\bea
v_1(t,x_1,x_2)=\frac{x_1}{L}v_{e}+
\frac{\hbar}{m}\mathfrak{Im}\left(\frac{\sum_{n}e^{i\phi(n)}e^{-i\frac{E(n)\tau(t)}{\hbar}}\frac{n_1(n)\pi}{L}\cos{\frac{n_1(n)\pi x_1}{L}}\sin{\frac{n_2(n)\pi x_2}{L}}}{\sum_{l}e^{i\phi(l)}e^{-i\frac{E(l)\tau(t)}{\hbar}}\sin{\frac{n_1(l)\pi x_1}{L}}\sin{\frac{n_2(l)\pi x_2}{L}}}\right)\nonumber\\
v_2(t,x_1,x_2)=\frac{x_2}{L}v_{e}+
\frac{\hbar}{m}\mathfrak{Im}\left(\frac{\sum_{n}e^{i\phi(n)}e^{-i\frac{E(n)\tau(t)}{\hbar}}\frac{n_2(n)\pi}{L}\sin{\frac{n_1(n)\pi x_1}{L}}\cos{\frac{n_2(n)\pi x_2}{L}}}{\sum_{l}e^{i\phi(l)}e^{-i\frac{E(l)\tau(t)}{\hbar}}\sin{\frac{n_1(l)\pi x_1}{L}}\sin{\frac{n_2(l)\pi x_2}{L}}}\right)\label{bvel}
\eea
\end{widetext}
where $E(k)=\frac{\hbar^2\pi^2(n^2_1(k)+n^2_2(k))}{2mL_0^2}$. 
We can expect that the dynamical behavior (i.e. the type of the trajectories) will depend on the ratio between $v_e$ and $\frac{\hbar}{mL}$. This is illustrated in Fig. (\ref{fig:fig1}).
\subsection{Quantum non-equilibrium}
In the pilot-wave theory, quantum ensembles in which the Born Law is violated
\be
\rho(t,\vec{x})\neq|\psi(t,\vec{x})|^2
\ee
are in principle allowed. However, we don't see such ensembles around us today as every experiment is in agreement with Born's Law. 
Hence if non-equilibrium existed in the very early universe it had to relax quickly to quantum equilibrium, apart from a possible persistence in exotic systems.
For this scenario to hold, it is necessary that relaxation occurs naturally, at least when the wave-function contains sufficiently many modes.

Since the quantity $\rho(t,\vec{x})/|\psi(t,\vec{x})|^2$ is conserved along a trajectory, there can never be relaxation without any further assumption. Indeed, 
if $\vec{x}^{-1}_{t\rightarrow t_i}(\vec{x})$ denotes the initial position which, when evolved from $t_i$ to $t$, returns $\vec{x}$, then we have that 
\be
\rho(t,{\vec{x}})=|\psi(t,\vec{x})|^2\frac{\rho(t_i,\vec{x}^{-1}_{t\rightarrow t_i}(\vec{x}))}{|\psi(t_i,\vec{x}^{-1}_{t\rightarrow t_i}(\vec{x}))|^2}\,.\label{liou}
\ee
Thus $\rho(t,\vec{x})$ can only be equal to $|\psi(t,\vec{x})|^2$ if both distributions are also identical initially. 
If relaxation takes place, it must therefore be on a coarse-grained level. 
In order to define the coarse-grained distributions, the domain of interest is divided in non-overlapping square cells of side length $\epsilon$ which are referred to as coarse-graining cells (CG cells for short).
Then the coarse-grained distributions are introduced as 
\be
\bar{r}(t,\vec{x})=\frac{1}{\epsilon^2}\int_{\mathrm{CG\,cell}\in \vec{x}}r(t,\vec{u}) du_1 du_2\,,
\ee
where $r$ is either $\rho$ or $|\psi|^2$, together with the coarse-grained H-function
\be
\bar{H}(t)=\int \bar{\rho}(t,\vec{x})\log{\frac{\bar{\rho}(t,\vec{x})}{\overline{|\psi(t,\vec{x})|^2}}}dx_1 dx_2\label{cgh}
\ee
which measures the difference between the two distributions. 
In \cite{ava1}, it is shown that $\frac{d \bar{H}(t)}{dt}\leq 0$ under certain assumptions about the initial distributions, indicating a tendency to relax to quantum equilibrium.
However the inequality is not strict and compatible with the existence of residues. 
Since then, relaxation to quantum equilibrium has been confirmed by many numerical simulations. 

\begin{figure}
\centering
\includegraphics[width=0.45\textwidth]{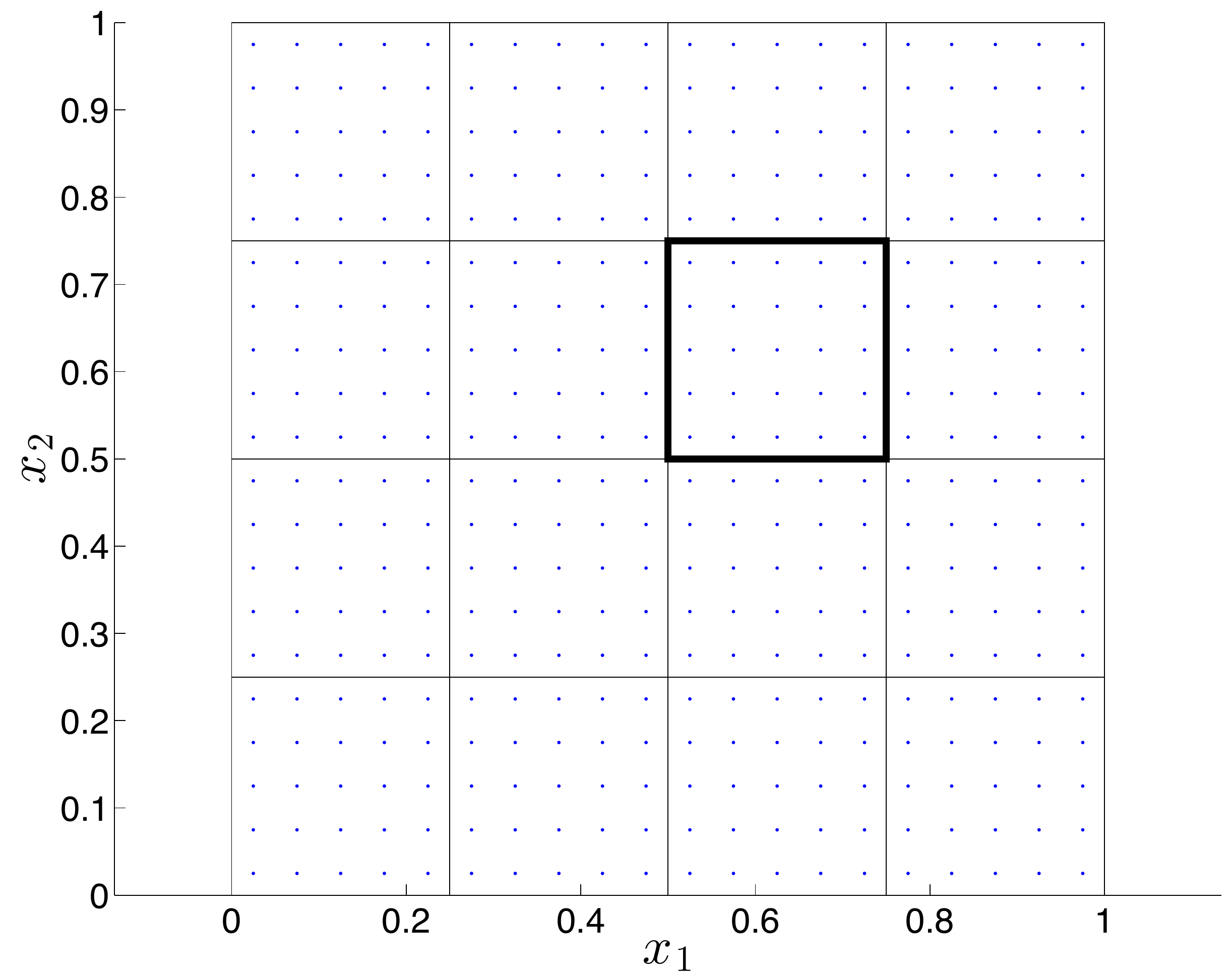}
\caption{\label{fig:fig2}Example of a decomposition of a fixed box of unit length used in the estimation of the coarse-grained H-function. 
In this case, we have that $C=4$, $D=5$ and $K=20$.
}
\end{figure}

The algorithm for the numerical estimation of $\bar{H}(t)$ used in \cite{vawe} is as follows. The fixed box is divided in $C^2$ CG cells.
A square grid of $K^2$ lattice points is then defined, with $K$ a multiple of $C$ ($K=C D$). 
The positions of the lattice points are  
\be
\vec{x}_{k,l}=(k\delta-0.5\delta,l\delta-0.5\delta)\quad\mathrm{with}\,k,l\in\{1,2,\ldots,K\}
\ee
and $\delta=L/K$ (see Fig. \ref{fig:fig2} for an example).  
The estimate of the CG H-function is defined as 
\be
\bar{h}(t)=\sum_{a=1}^{a=C}\sum_{b=1}^{b=C}\bar{h}_{a,b}(t)=\epsilon^2\sum_{a=1}^{a=C}\sum_{b=1}^{b=C}\bar{\rho}_{a,b}(t)\log{\frac{\bar{\rho}_{a,b}(t)}{\overline{|\psi|^2}_{a,b}(t)}}\,.\label{hb}
\ee
The value of $\bar{\rho}_{a,b}(t)$ is obtained by averaging the values of $\rho(t,\vec{x}_{lat})$ over the lattice points $\vec{x}_{lat}$ contained 
within the cell $(a,b)$. Each value $\rho(t,\vec{x}_{lat})$ is obtained thanks to Eq. (\ref{liou}).

The previous algorithm is the back-tracking algorithm. We can also define a forward-tracking algorithm.
We randomly generate $P$ positions according to the initial non-equilibrium density. We let them evolve, we count the number of positions in each CG cell $(a,b)$ at time $t$ and we divide it by $P$ 
in order to define $\bar{\rho}_{a,b}(t)$. Both methods have respective advantages.

In the present case, the novelty is that the domain is expanding. 
If we want to keep the coarse-graining length fixed, we would have to record the values of $\bar{H}(t)$ at times $t_n=n\frac{\epsilon}{v_e}$, $n$ being a positive integer. 
Only for those time values do we have that the side length $L(t)$ is a multiple of the coarse-graining length $\epsilon$.
Another possibility is to increase the coarse-graining length with time (then we would use $\epsilon s(t)$ instead of $\epsilon$) -- 
while this makes sense for a field on an expanding space, in which case the expansion is a global mechanism affecting everything, 
it does not seem to be appropriate for the present case, in which the coarse-graining length is related to the finite (non-expanding) size of a detector.
We will see our both definitions are related in a subsection \ref{subsec:expl}.
\subsection{Examples}
For the following three examples, we use the following parameters: $m=10^{-30}\,\mathrm{kg}$, $L_0=1\,\mathrm{m}$, $\epsilon=0.05\,\mathrm{m}$ and $\mathrm{v_e}=1\,\mathrm{m}\,\mathrm{s}^{-1}$. 
Therefore there are $20\times20$ CG cells initially. What changes from one example to the other is the initial non-equilibrium distribution.

We estimate the coarse-grained H-function at the times $t_n=0.05 n$, for $n$ up to $20$, time by which the side length has doubled.
The values $\bar{h}(t_n)$ are computed using the two methods, backward and forward tracking (with $D=32$ and $P=4\times 10^6$). We also plot the evolution of 
\be
\bar{g}(t_n)=\epsilon^2\sum_{a=1}^{C(t_n)}\sum_{b=1}^{C(t_n)}|\bar{\rho}_{a,b}(t_n)-\overline{|\psi|^2}_{a,b}(t_n)|\label{gb}
\ee
which is another measure of the difference between $\bar{\rho}$ and $\overline{|\psi|^2}$, used for instance in \cite{efthy0,efthy,hawicodu}.
\begin{figure}
\centering
\includegraphics[width=0.45\textwidth]{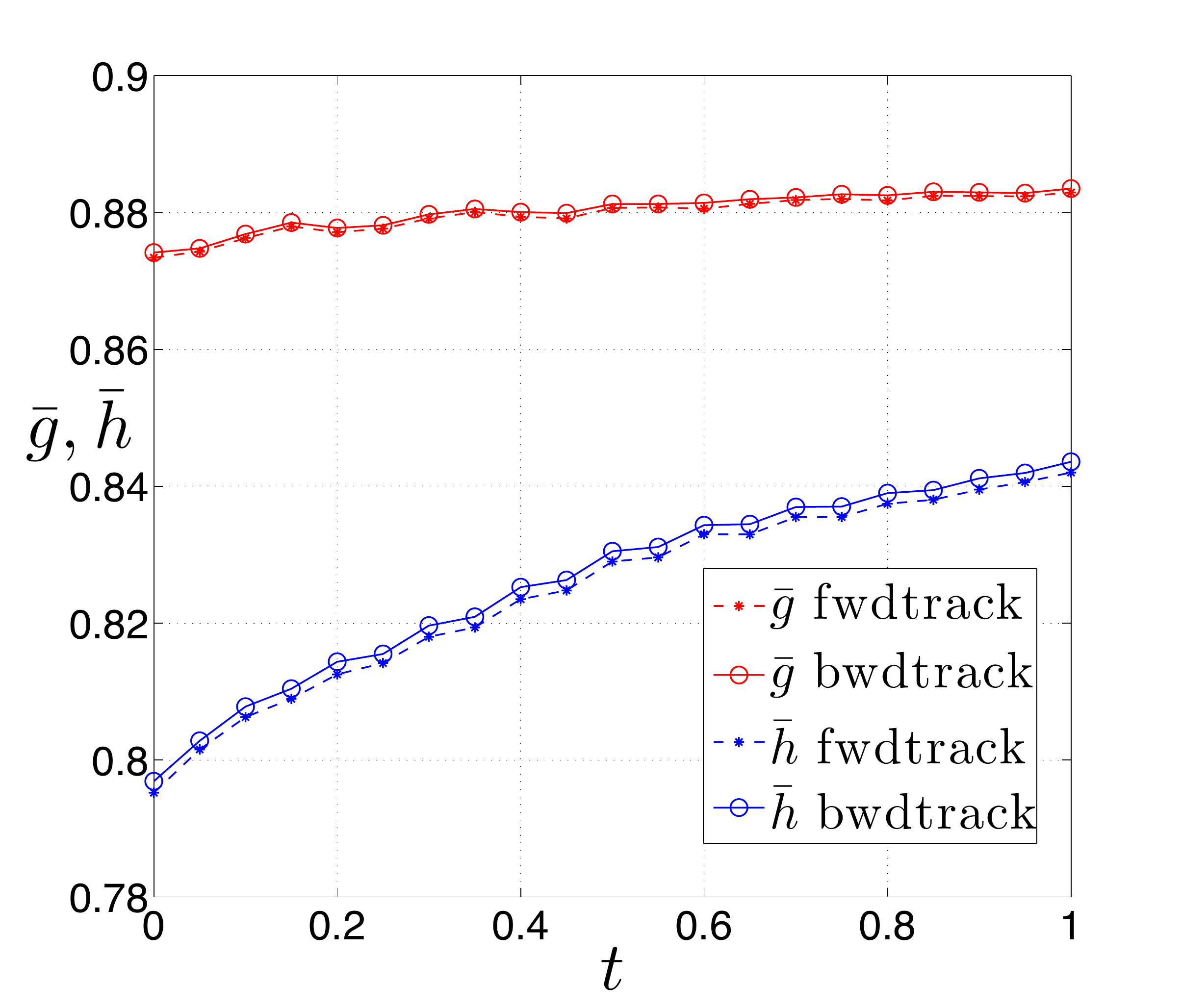}
\caption{\label{fig:fig3}Plot of $\bar{h}$ (Eq. (\ref{hb})) and $\bar{g}$ (Eq. (\ref{gb})) for $\rho_0$ (Eq. (\ref{rho0})).}
\end{figure}
We also introduce the mean relative difference between $\bar{\rho}$ and $\overline{|\psi|^2}$ as
\bea
\bar{f}(t_n)=\frac{1}{C^2(t_n)}\sum_{a=1}^{C(t_n)}\sum_{b=1}^{C(t_n)}\frac{|\bar{\rho}_{a,b}(t_n)-\overline{|\psi|^2}_{a,b}(t_n)|}{\overline{|\psi|^2}_{a,b}(t_n)}.\nonumber\\\label{fb}
\eea
\begin{figure}
\centering
\includegraphics[width=0.45\textwidth]{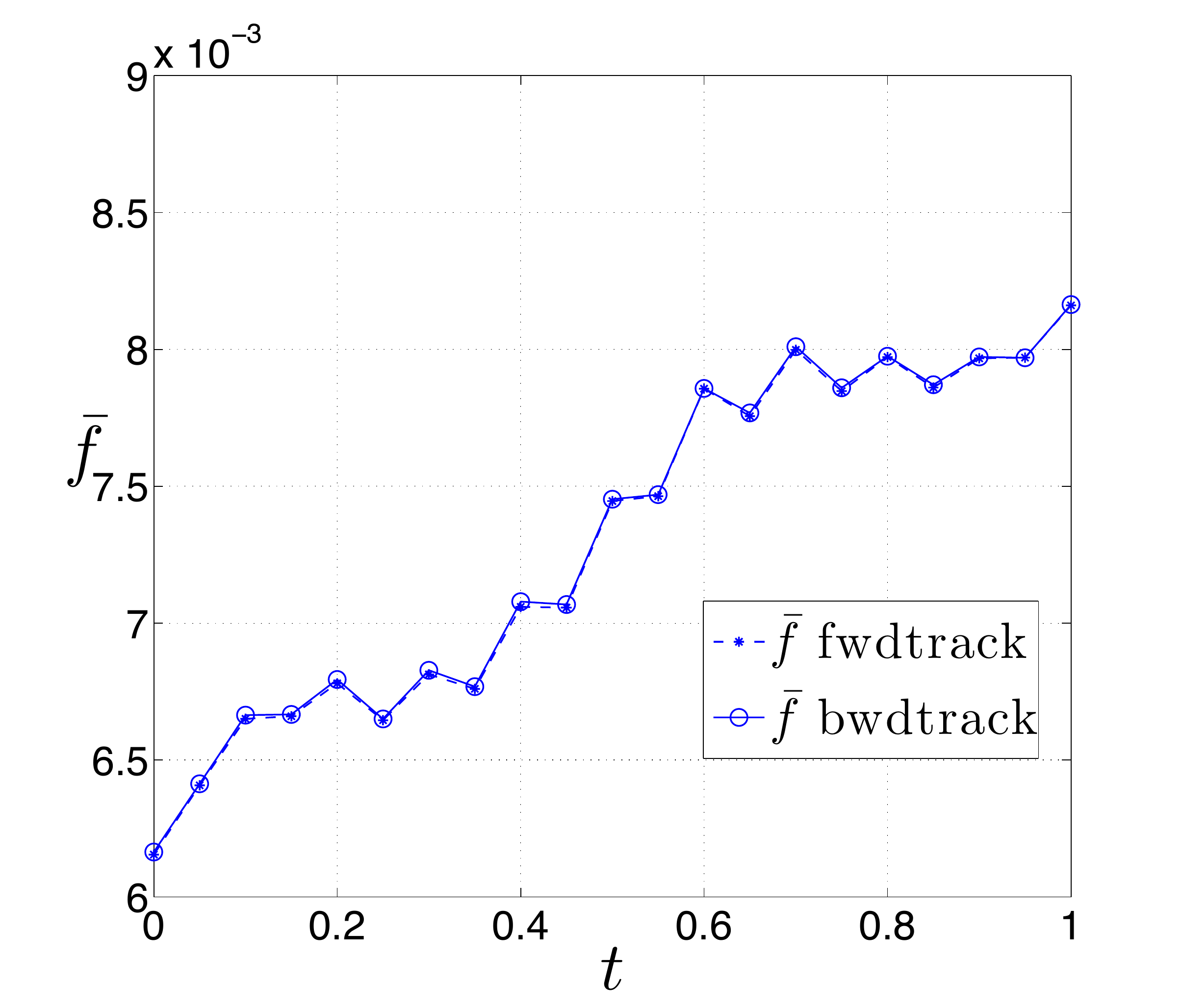}
\caption{\label{fig:fig4}Plot of $\bar{f}$ (Eq. (\ref{fb})) for $\rho_0$ (Eq. (\ref{rho0})).}
\end{figure}
\subsubsection{Example 1}
The initial non-equilibrium density is 
\be
\rho_0(t_0,\vec{x})=\frac{4}{L_0^2}\sin^2(\frac{\pi x_1}{L_0})\sin^2(\frac{\pi x_2}{L_0})\,,\label{rho0}
\ee
which is significantly different from $|\psi(t_0,\vec{x})|^2$.
The evolutions of $\bar{h}(t_n)$ and $\bar{g}(t_n)$ can be visualized in Fig. \ref{fig:fig3} and that of $\bar{f}(t_n)$ in Fig. \ref{fig:fig4}. 
At the final time there is an increase of about $5\%$ for the estimate of the standard coarse-grained H-function.
\subsubsection{Example 2}
\begin{figure}
\centering
\includegraphics[width=0.45\textwidth]{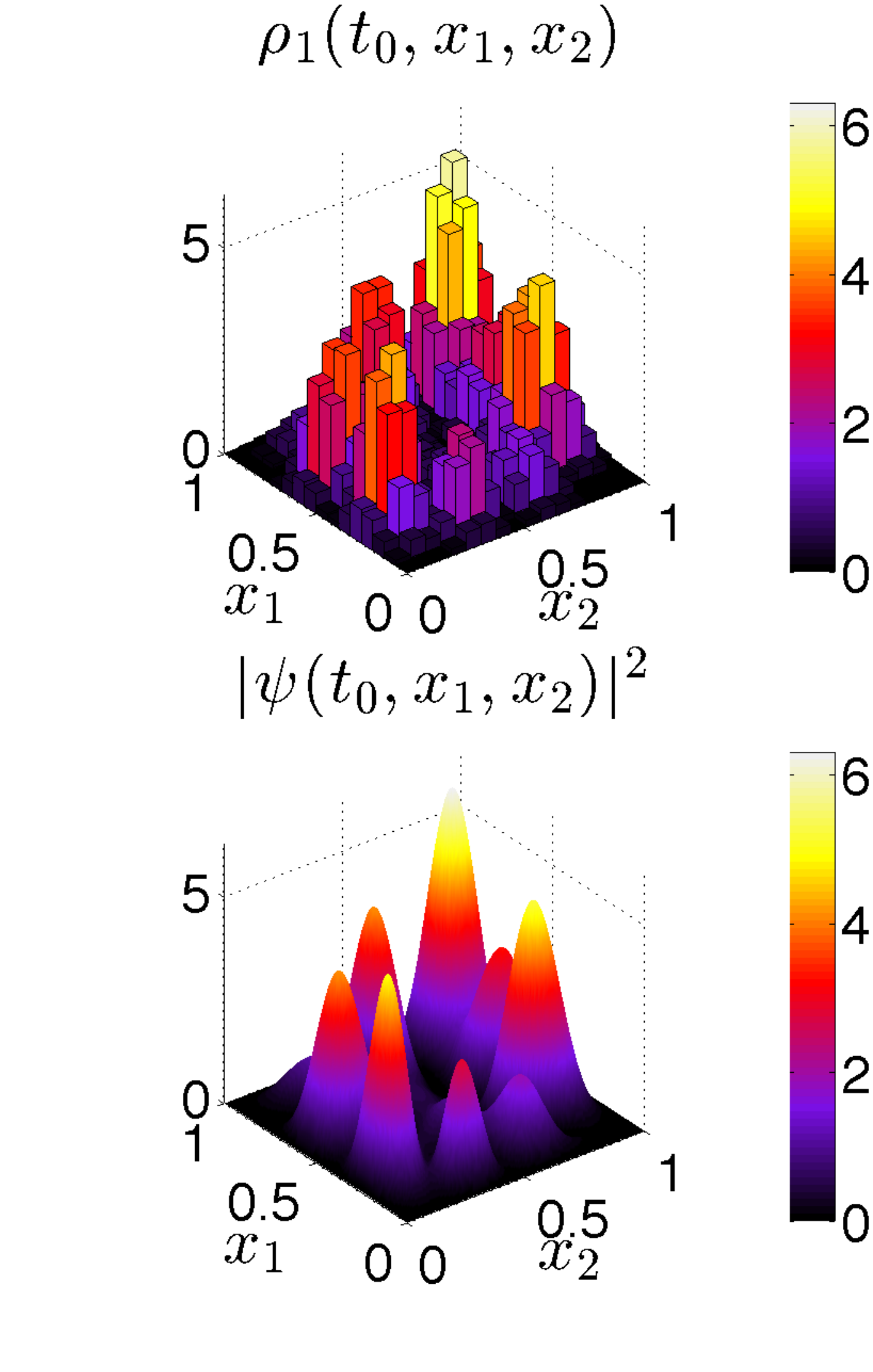}
\caption{\label{fig:fig5}Initial non-equilibrium distribution $\rho_1(t_0,x_1,x_2)$ and quantum equilibrium distribution $|\psi(t_0,x_1,x_2)|^2$.}
\end{figure}

\begin{figure}
\centering
\includegraphics[width=0.45\textwidth]{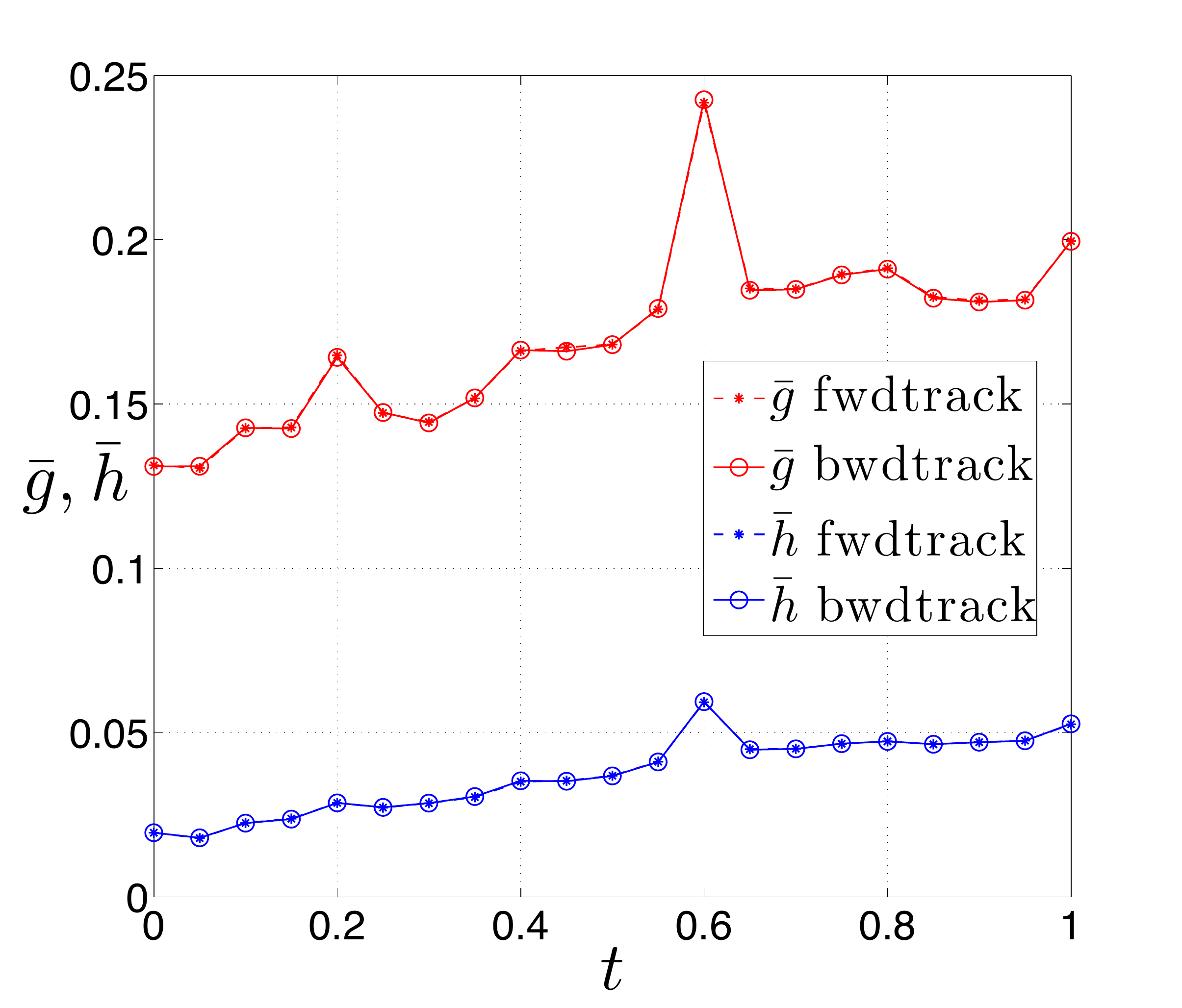}
\caption{\label{fig:fig6}Plot of $\bar{h}$ (Eq. (\ref{hb})) and $\bar{g}$ (Eq. (\ref{gb})) for $\rho_1$ (see Fig. \ref{fig:fig5}).}
\end{figure}

We use the same parameters as in the first example but we choose an initial non-equilibrium distribution $\rho_1$ which is not far from quantum-equilibrium. 
To obtain that distribution, we have performed a coarse-graining on the quantum equilibrium distribution with a length $\epsilon=\frac{1}{16}\,\mathrm{m}$. 
Both distributions can be visualized in Fig. \ref{fig:fig5}.
The evolution of the functions $\bar{h}$  and $\bar{g}$ is plotted in Fig. \ref{fig:fig6}. 
Again there is a global increase in all functions measuring the distance from quantum equilibrium.

\begin{figure}
\centering
\includegraphics[width=0.45\textwidth]{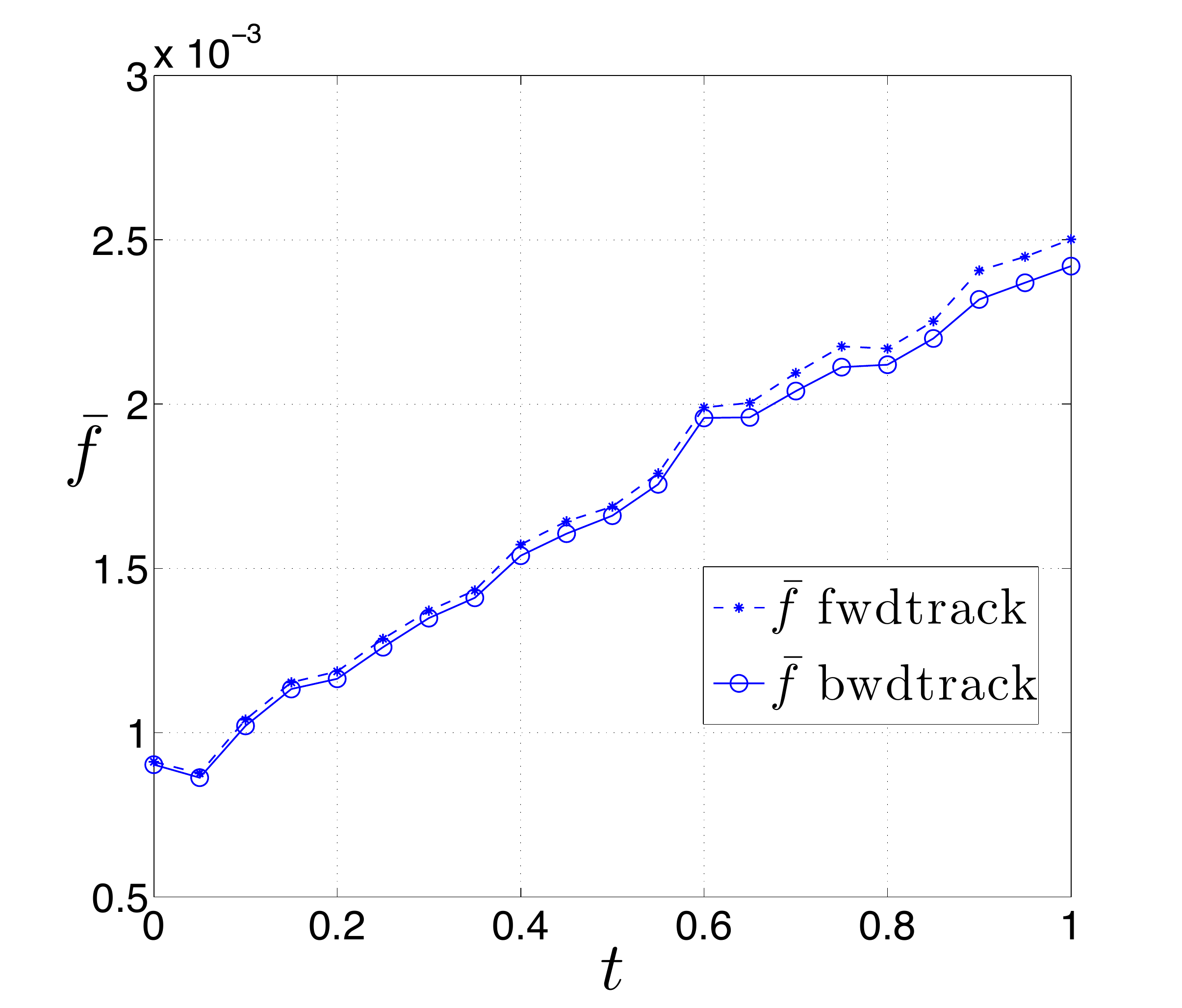}
\caption{\label{fig:fig7}Plot of $\bar{f}$ (Eq. (\ref{fb}))  for $\rho_1$ (see Fig. \ref{fig:fig5}).}
\end{figure}  

\subsubsection{Example 3}
\begin{figure}
\centering
\includegraphics[width=0.45\textwidth]{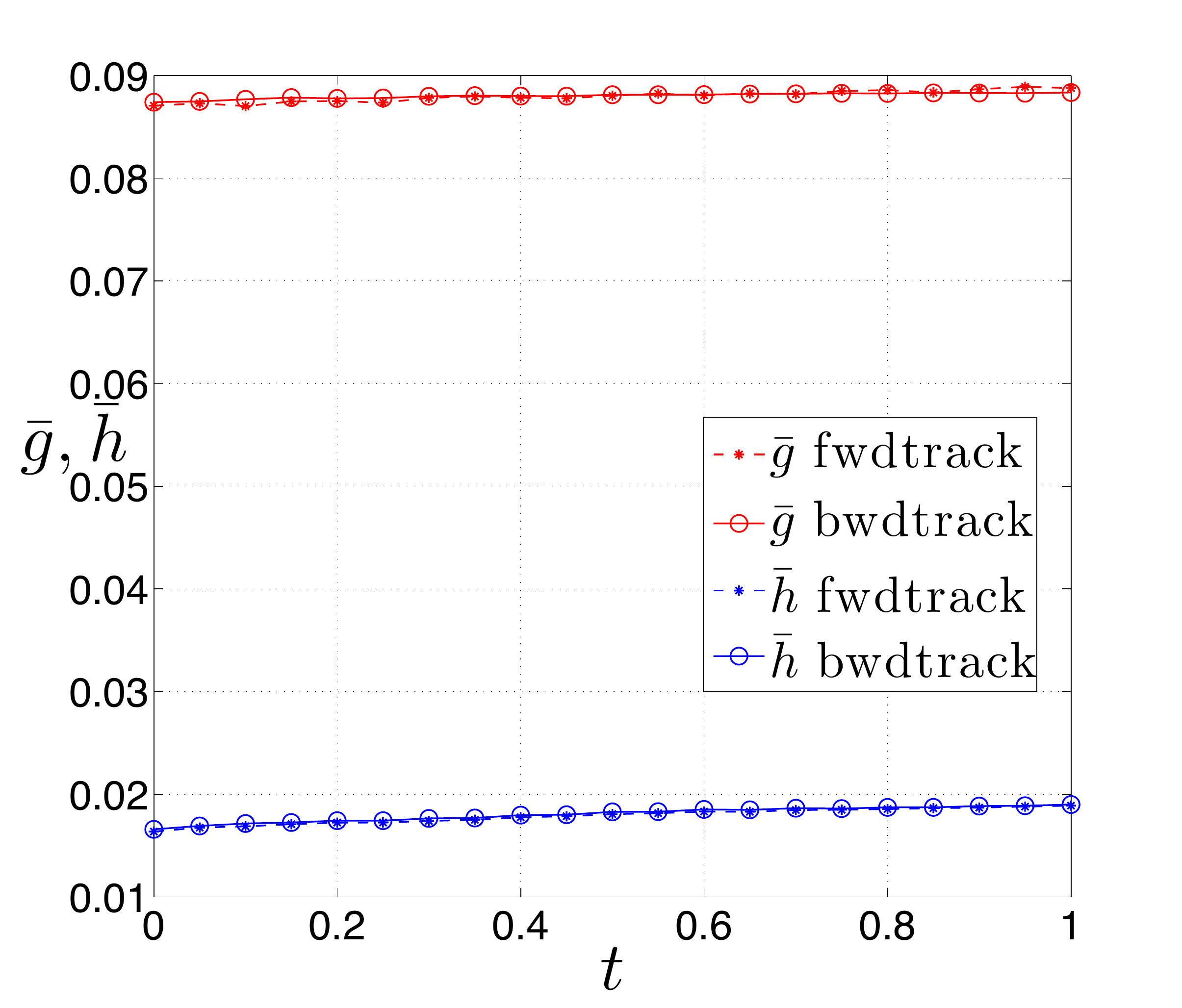}
\caption{\label{fig:fig8}Plot of $\bar{h}$ (Eq. (\ref{hb})) and $\bar{g}$ (Eq. (\ref{gb})) for $\rho_2$.}
\end{figure}

\begin{figure}
\centering
\includegraphics[width=0.45\textwidth]{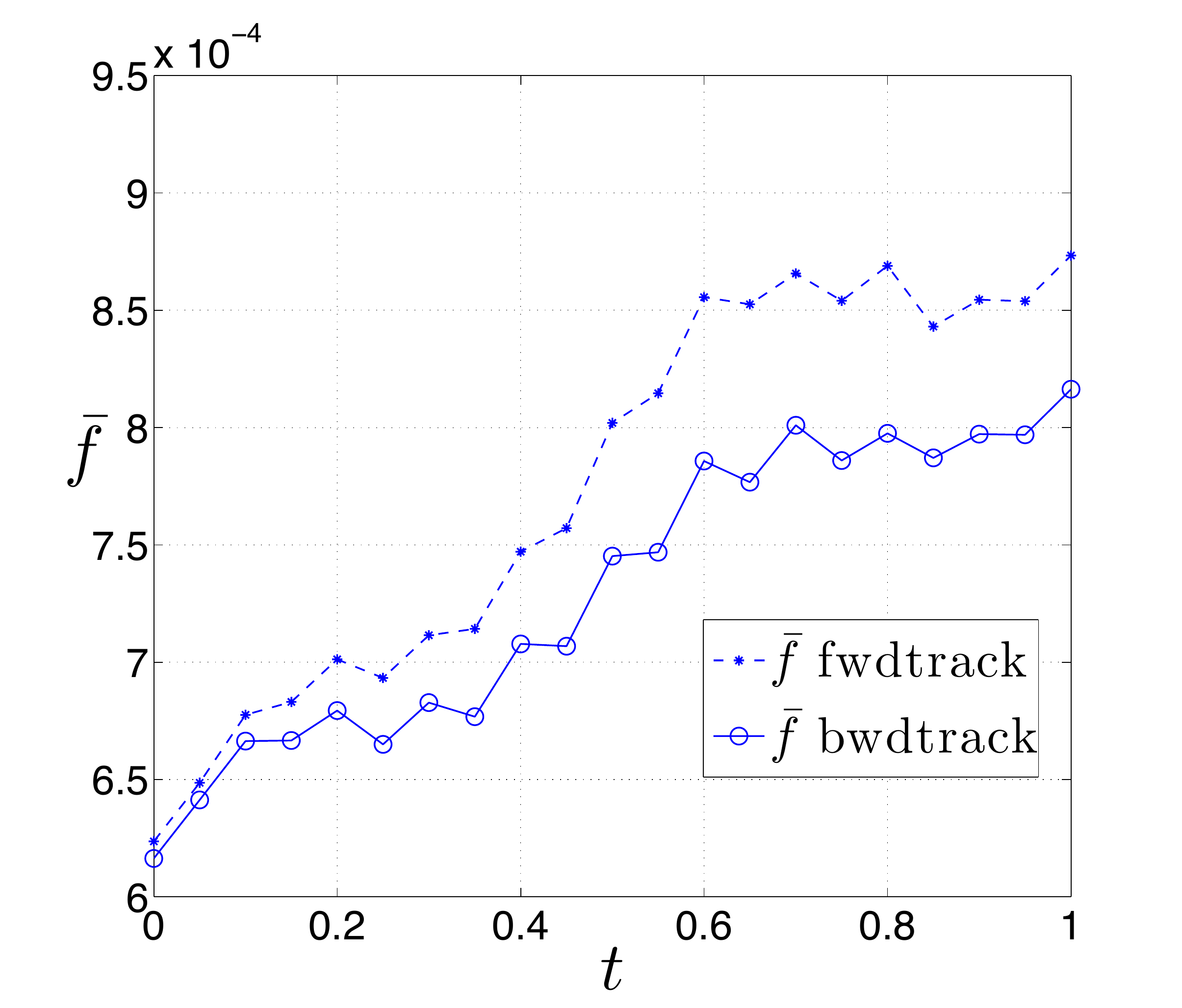}
\caption{\label{fig:fig9}Plot of $\bar{f}$ (Eq. (\ref{fb})) for $\rho_2$.}
\end{figure}

The initial non-equilibrium density is 
\be
\rho_2(t_0,\vec{x})=0.9|\psi(t_0,\vec{x})|^2+0.1\rho_0(t_0,\vec{x})\,.\label{rho2}
\ee
Again the distribution is close to equilibrium and we have a final increase comparable to that of the first example, as can been seen from Fig. (\ref{fig:fig8}) and Fig. (\ref{fig:fig9}).

\subsection{Explanation}\label{subsec:expl}
We introduce a second coarse-graining, with a time-dependent length $\epsilon(t)=s(t)\epsilon$; its action on a function $r$ is represented by the tilde symbol:
\be
\tilde{r}(t,\vec{x})=\frac{1}{\epsilon^2(t)}\displaystyle\int_{\mathrm{CG\,cell(t)}\in \vec{x}}r(t,\vec{u}) du_1 du_2\,,
\ee
where $\mathrm{CG\,cell(t)}$ is a CG cell of side length $\epsilon(t)$. The coarse-grained H-function with time-dependent coarse-graining length is 
\be
\tilde{H}(t)=\int \tilde{\rho}(t,\vec{x})\log{\frac{\tilde{\rho}(t,\vec{x})}{\widetilde{|\psi(t,\vec{x})|^2}}}dx_1 dx_2\label{cgh}\,.
\ee
One can show that the $\tilde{H}(t)=\bar{H}'(\tau(t))$, where prime refers to a second system, describing a particle in a 2D square box of fixed length $L_0$. 
The wave-function in this second system is the same as the one in the first system (see Eq. (\ref{wf}) for the generic form), except that the solutions $\psi_{n_1,n_2}$ are replaced by normalized energy eigenstates, and $\vec{x}$ by $\vec{y}$.
To proof this, we consider the rescaled trajectory in the first system, from $t_0$ up to $t_f$, and we show that it is equivalent to a trajectory in the second system evolving from $t_0$ up to $\tau(t_f)$ (the initial condition is the same in both systems); 
we start from $\vec{y}(t)=\vec{x}(t)/s(t)$, we take its time-derivate and we use the explicit expression for the velocity-field (Eq. (\ref{bvel})), then the rest follows after a change of time variable. 
(There is a similar derivation in the case of a scalar field on an expanding space in \cite{cova1}).

In Fig. \ref{fig:fig8}, we have plotted the time $t$ versus $\tau(t)$. We see that $\tau(t)<t$ for $t>t_0$, hence we call $\tau(t)$ a retarded time; moreover $\tau(t)$ quickly approaches an asymptotical value (1 in this case). 
As far as relaxation goes, as measured with an increasing coarse-graining length $\epsilon(t)$, it will quickly get stopped or ``frozen''.

The last step in the explanation is to show that \mbox{$\tilde{H}(t)<\bar{H}(t)$} for $t>0$. 
We don't have a general proof yet but it seems plausible; indeed, as the coarse-graining length increases, it will finally reach the length of the domain for which we have that $\tilde{H}(t)=0$.
\begin{figure}
\centering
\includegraphics[width=0.45\textwidth]{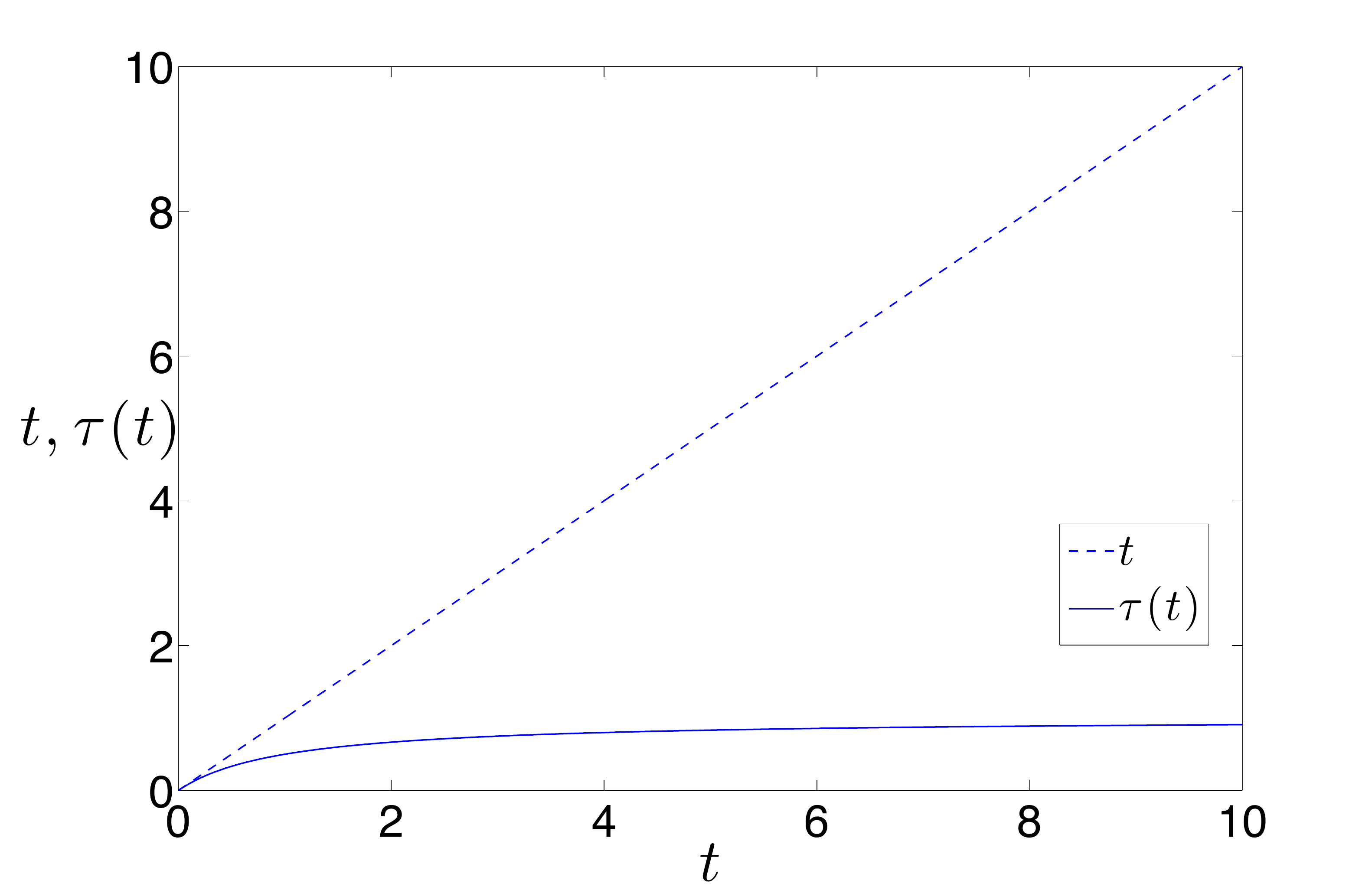}
\caption{\label{fig:fig10}Plot of $t$ versus $\tau(t)=\frac{L_0 t}{L_t}$, for $L_0=1\,\mathrm{m}$ and $v_e=1\,\mathrm{m}\,\mathrm{s}^{-1}$.}
\end{figure}

\section{Conclusion}\label{sec:dis}

Let us assume that the universe is governed by pilot-wave dynamics and that it was in quantum non-equilibrium at some very early stage (at the subsystem level).
The most natural candidate for its detection today would be exotic relic particles. 
Another possibility is that quantum non-equilibrium is generated in more exotic systems, perhaps at the interface between quantum mechanics and gravity.
Therefore the idea of quantum non-equilibrium can seem far-fetched, especially since quantum equilibrium is a terminal state.
 
In the case of quantum fields on an expanding space, it was already understood that the expansion acts again the natural process of relaxation, without preventing it however, only slowing it down \cite{ava3,cova1}.
Here we have shown examples of expanding quantum ensembles that move away from quantum equilibrium, although they start close to quantum equilibrium 
(in the case of the field on expanding space, the coarse-graining length increases with time, in the case of the confined particle it doesn't, hence the difference in behavior).
It is interesting because now we can imagine that an ensemble with a very small residual non-equilibrium could have its value amplifed.
This would be achieved thanks to an experimental protocol based on trapping and expansion. 
In future work, we plan to extend the case that we have considered to the relativistic realm (by considering Dirac particles), and to make such an experimental protocol more pragmatic.

Related ideas haven been developed in section 7 of \cite{avajpa}, in which it is pointed out that ensembles of particles emitted from astrophysical sources could have their non-equilibrium microscopic scales 
stretched to macroscopic ones through the spreading of their wave packets.
Our work is different in the sense that we are not concerned with the history of an ensemble from its emission to its detection (and the inherent difficulties described in \cite{avajpa}). 
We assume that we have this ensemble at our diposal and that it is almost in quantum equilibrium.
We hope to have bring some support to the idea that it can be pushed further away from quantum equilibrium.
\begin{acknowledgments}
I thank Alex Matzkin and Antony Valentini for their comments on an earlier manuscript.
\end{acknowledgments}

\bibliography{refs}
\appendix
\section{Wave-function parameters}
The wave-function used for the simulations is 
\be
\psi(t,x_1,x_2)=\sum_{n=1}^{n=10}\frac{e^{i\phi(n)}}{\sqrt{10}}\psi_{n_1(n),n_2(n)}(t,x_1,x_2)
\ee
where $\psi_{n_1(n),n_2(n)}(t,x_1,x_2)$ is defined at Eq. (\ref{mode}) and where the quantum numbers and phase angles can be found in the following table. 
\begin{center}
\begin{tabular}{|c|c|c|c|}
\hline
$n$ & $n_1(n)$ & $n_2(n)$ & $\phi_{n}/{2\pi}$\\
\hline
$1$ & $1$ & $1$& $0.5007885937046778$\\
$2$ & $1$ & $2$& $0.2563559569433025$\\
$3$ & $2$ & $1$& $0.0577194737040234$\\
$4$ & $2$ & $2$& $0.5942444602612857$\\
$5$ & $1$ & $3$& $0.9461819879073565$\\
$6$ & $3$ & $1$& $0.5466682505848018$\\
$7$ & $2$ & $3$& $0.1652644360494799$\\
$8$ & $3$ & $2$& $0.3915951186360821$\\
$9$ & $1$ & $4$& $0.9067195609839858$\\
$10$ & $4$ & $1$& $0.4541288770927727$\\
\hline
\end{tabular}
\end{center}
\end{document}